\documentclass[12pt]{article}
\usepackage{graphics,color}
\begin{document}
\thispagestyle{empty}
\noindent\
\\
\\
\\
\begin{center}
\large \bf Flavor Mixing and Neutrino Masses
\end{center}
\hfill
 \vspace*{1cm}
\noindent
\begin{center}
{\bf Harald Fritzsch}\\
Department f\"ur Physik\\
Universit\"at M\"unchen\\
Theresienstra{\ss}e 37\\
D-80333 M\"unchen,
Germany
\vspace*{0.5cm}
\end{center}

\begin{abstract}

We discuss mass matrices with four texture zeros for the quarks and leptons. The three mixing angles for the quarks and leptons are functions of the fermion masses. The results agree with the experimental data. The ratio of the masses of the first
two neutrinos is given by the solar mixing angle. The neutrino masses are calculated:\\ $m_1$ $\approx$ 0.004 eV , $m_2$ $\approx$ 0.010 eV , $m_3$ $\approx$ 0.070 eV.

\end{abstract}

\newpage

-----------------------------Flavor Mixing of Quarks----------------------------\\

The flavor mixing of the quarks is parametrized by the CKM-matrix.
There are several ways to describe the CKM-matrix in terms of three
angles and one phase parameter. I prefer the parametrization, which Z. Xing and I introduced years ago (ref.(1)), given by the angles $\theta_u$,
$\theta_d$ and $\theta$:
\begin{eqnarray}
U = \left( \matrix{ c^{}_{u}
& s^{}_{u} & 0 \cr
-s^{}_{u} &
c^{}_{u} &
0 \cr 0 & 0 & 1 \cr} \right)\times \left( \matrix{ e^{-i\phi}
& 0 & 0 \cr
0 &
c &
s \cr 0 & -s & c \cr} \right) \times \left( \matrix{ c^{}_{d}
& - s^{}_{d} & 0 \cr
s^{}_{d} &
c^{}_{d} &
0 \cr 0 & 0 & 1 \cr} \right),
\end{eqnarray}
\\
$c^{}_{u,d} \sim \cos\theta^{}_{u,d}$, $s^{~}_{u,d}
\sim \sin\theta^{}_{u,d}$, $c \sim \cos\theta$, $s \sim
\sin\theta$.\\

The three angles have been determined by the experiments:\\

$\theta^{}_u \simeq 5.4^\circ$, $\theta^{}_d \simeq 11.7^\circ$, $\theta^{}\simeq 2.4^\circ$.\\

Relations between the quark masses and the mixing angles can be derived, if the quark mass matrices have "texture zeros", as shown by S. Weinberg and me in 1977 (ref.(2)). For six quarks the mass matrices have four "texture zeros":\\
\begin{eqnarray}
M= \left( \matrix{ 0 & A & 0 \cr A^* & 0 & B \cr
0 & B^* & C \cr} \right) \;.
\end{eqnarray}
\\
We can calculate the angles $\theta_u$ and
$\theta_d$ as functions of the mass eigenvalues:
\\
\begin{equation}
\theta^{}_u \simeq
\sqrt{m^{}_u/m^{}_c},\hspace*{1cm}
\theta^{}_d \simeq
\sqrt{m^{}_d/m^{}_s}.
\end{equation}
\\
Using the observed masses for the quarks, we find for these angles:\\

$\theta^{}_d \simeq (13.0\pm 0.4)^\circ,\hspace*{1cm} \theta^{}_u \simeq (5.0\pm 0.7)^\circ.$\\
\\
The experimental values agree with the theoretical results:\\

$\theta^{}_d \simeq (11.7\pm 2.6)^\circ,\hspace*{1cm} \theta^{}_u \simeq (5.4\pm 1.1)^\circ$.\\
\\
We can also calculate the Cabibbo angle:
\\
\begin{equation}
\theta^{}_c  \cong|\sqrt{m^{}_d/m^{}_s}+e^{i\phi}\sqrt{m^{}_u/m^{}_c}|.
\end{equation}
\\
Taking into account the observed quark masses, the phase angle $\phi$ must be close to $90^\circ$.\\

The CKM-element $V_{cb}$ has been determined by the experiments:
\\
\begin{equation}
V_{cb} \cong 0.041.
\end{equation}
\\
The "texture zeros" imply a relation between the four heavy quark masses, a phase angle $\psi$ and $V_{cb}$:
\\
\begin{equation}
V_{cb} \cong|\sqrt{m^{}_s/m^{}_b}+e^{i\psi}\sqrt{m^{}_c/m^{}_t}|.
\end{equation}
\\
This relation seems to be a problem, if the mass of the top quark is
as large as 174 GeV. For example, let us assume that the phase angle is $180^\circ$ and let us take the following values for the quark masses:
$m_s\simeq 0.11 ~GeV, ~m_b\simeq 4.3 ~GeV, ~m_c\simeq 1.1 ~GeV, ~m_t\simeq 174 ~GeV$. The result
disagrees with the experimental result:
\begin{equation}
V_{cb} \cong \sqrt{m^{}_s/m^{}_b} -
\sqrt{m^{}_c/m^{}_t} \cong 0.08.
\end{equation}

The "texture zero" - matrices are related to specific symmetries, which are present at the
energy scale of the Grand Unification. But the flavor mixing angles and the quark masses are
measured at relatively low energies. The observed quark masses receive also a small contribution
from electromagnetic radiative corrections, which are proportional to the fine structure constant.
An observed quark mass is the sum of a "bare quark mass", given by the "texture zero" mass matrix,
and a radiative correction. The relations
between the flavor mixing angles and the quark masses are relations, involving the bare quark masses - they do not include the radiative corrections.\\

Let us consider relation (6). The observed value of the CKM element $V_{cb}$ is about 0.041. Taking
into account radiative corrections, the two terms in relation (6) vary as follows:

 \begin{eqnarray}
 0.11 \le\sqrt{m^{}_s/m^{}_b}\le 0.21, \hspace*{1cm}
 0.06 \le\sqrt{m^{}_c/m^{}_t}\le 0.10.
\end{eqnarray}
\\
Thus for certain values of the bare quark masses the relation (6) is valid, e.g. for the values:
$m_t\simeq 172 ~GeV$,~$m_c\simeq 1.5 ~GeV$,~$m_b\simeq 4.2 ~GeV$,~$m_s\simeq 0.08 ~GeV$.\\

The unitarity triangle is the triangle, which is obtained in the complex plane, if
the first column and the complex conjugated third column of the CKM matrix are multiplied. This
scalar product must be zero, since the CKM matrix is a unitary matrix.\\

If the flavor mixing angles are given by the quark mass ratios, the sides of the triangle are determined by the following three numbers:
\\
\begin{equation}
\theta^{}_u \simeq
\sqrt{m^{}_u/m^{}_c},\hspace*{1cm}
\theta^{}_d \simeq
\sqrt{m^{}_d/m^{}_s},\hspace*{1cm}
\theta^{}_c  \cong|\sqrt{m^{}_d/m^{}_s}+e^{i\phi}\sqrt{m^{}_u/m^{}_c}|.
\end{equation}
\\
The three angles of the unitarity triangles are usually denoted by $\alpha$, $\beta$ and $\gamma$. The
angle $\alpha$ is given by the angle $\phi$, introduced in eq.(4). It must be close to 90 degrees, in agreement
with the experiment: $\alpha \simeq 87^\circ...94^\circ$.\\

The CKM matrix can now be written as follows:
\begin{eqnarray}
U = \left( \matrix{ c^{}_{u}
& s^{}_{u} & 0 \cr
-s^{}_{u} &
c^{}_{u} &
0 \cr 0 & 0 & 1 \cr} \right)\times \left( \matrix{ i
& 0 & 0 \cr
0 &
c &
s \cr 0 & -s & c \cr} \right) \times \left( \matrix{ c^{}_{d}
& - s^{}_{d} & 0 \cr
s^{}_{d} &
c^{}_{d} &
0 \cr 0 & 0 & 1 \cr} \right).
\end{eqnarray}
This case might be described as "Maximal CP-Violation". Taking into account the observed angles, one finds the following mixing matrix:
\begin{eqnarray}
U \simeq \left( \matrix{ 0.97 i+0.02 & -0.20 i + 0.09  & 0.004 \cr -0.09 i + 0.02 & 0.02 i + 0.97 & 0.04 \cr
- 0.01 & - 0.04 & 1 \cr} \right) \;.
\end{eqnarray}
We mention another interesting feature of the quark masses:
\begin{eqnarray}
m(c) : m(u) \simeq m(t) : m(c) \simeq 207,\nonumber \\
m(s) : m(d) \simeq m(b) : m(s) \simeq 23.
\end{eqnarray}

Thus the logarithms of the quark masses (u,c,t) and (d,s,b) describe straight lines. Nobody understands, why
this is the case. I used this feature to predict in 1987 the mass of the t-quark: 175 GeV. Eight years later the t-quark with a mass of 173 GeV was discovered!\\
\\

----------------------------Flavor Mixing of Leptons----------------------------\\
\\

The flavor mixing of leptons is described by a $3\times 3$ unitary matrix $U$, similar to the CKM mixing matrix
for the quarks. It can be parametrized in terms of three angles and three phases. I use
a parametrization, introduced by Z. Xing and me (ref.(3)):
\begin{eqnarray}
U = \left( \matrix{ c^{}_{l}
& s^{}_{l} & 0 \cr
-s^{}_{l} &
c^{}_{l} &
0 \cr 0 & 0 & 1 \cr} \right)\times \left( \matrix{ e^{-i\phi}
& 0 & 0 \cr
0 &
c &
s \cr 0 & -s & c \cr} \right) \times \left( \matrix{ c^{}_{\nu}
& - s^{}_{\nu} & 0 \cr
s^{}_{\nu} &
c^{}_{\nu} &
0 \cr 0 & 0 & 1 \cr} \right) \times P^{}_\nu \,
\end{eqnarray}

$c^{}_{l,\nu} \sim \cos\theta^{}_{l,\nu}$, $s^{~}_{l,\nu}
\sim \sin\theta^{}_{l,\nu}$, $c \sim \cos\theta$, $s \sim
\sin\theta$.\\

$\theta^{}_{\nu}$ = $\theta^{}_{sun}$: solar angle,

$\theta$ = $\theta^{}_{at}$: atmospheric angle,

$\theta^{}_{l}$: reactor angle.
\\
\\
The phase matrix $P^{}_\nu ={\rm
Diag}\{e^{i\rho}, e^{i\sigma}, 1\}$ is relevant only, if the neutrino masses are Majorana masses. The neutrino oscillations are described by the three angles, which have been measured: $\theta^{}_{sun}=\theta^{}_\nu \simeq 34^\circ$,
$\theta^{}_{at}=\theta\simeq 45^\circ $, $\theta^{}_l \simeq 13^\circ. $
\\
We assume that the mass matrices of the leptons also have four texture zeros:
\begin{eqnarray}
M= \left( \matrix{ 0 & A & 0 \cr A^* & 0 & C \cr
0 & C^* & D \cr} \right) \;.
\end{eqnarray}
\\
In this case we can calculate the solar and the reactor angle as functions of the lepton masses:
 \begin{equation}
\tan\theta^{}_l \simeq
\sqrt{m^{}_e/(m^{}_e+m^{}_\mu)}, \hspace*{1cm}
\tan\theta^{}_\nu \simeq
\sqrt{m^{}_1/(m^{}_1+m^{}_2)}.
\end{equation}
\\
From the solar mixing angle we obtain for the neutrino mass ratio:
\begin{equation}
{m^{}_1/m^{}_2} \simeq 0.44.
\end{equation}
We use this relation and the experimental results for the mass differences of the neutrinos to
determine the three neutrino masses (ref. (4)):
\begin{eqnarray}
{m^{}_1} \simeq 0.004~eV , \nonumber \\
{m^{}_2} \simeq 0.010~eV , \nonumber \\
{m^{}_3} \simeq 0.070~eV.
\end{eqnarray}

The atmospheric mixing angle is determined by the lepton masses:
\begin{equation}
sin ~\theta_{at} \cong| \sqrt{\frac{m_3}{m_2+ m_3}}\times \sqrt{\frac{m_{\mu}}{m_{\mu}+ m_{\tau}}}+e^{i\alpha} \sqrt{\frac{m_2}{m_2+ m_3}}\times \sqrt{\frac{m_{\tau}}{m_{\mu}+ m_{\tau}}}|
\end{equation}
Here the phase parameter $\alpha$ is not known. Thus the atmospheric angle should be in the
range
\begin{equation}
13^\circ< \theta_{at} < 40^\circ.
\end{equation}

If $\alpha$ vanishes, one has $\theta_{at} = 40^\circ $, in agreement with the experimental value.
It is interesting to observe that here the phase parameter must be close to zero, in eq. (6) it must be
close to $180^\circ $, and in eq. (4) close to $90^\circ $. Perhaps this follows from a specific symmetry
of the "texture zero" mass matrices.\\

The mass matrices of the quarks and leptons are not exactly given by
texture zero matrices. Radiative corrections of the
order of the fine-structure constant $\alpha$ will contribute - the zeros will be replaced by small numbers.\\

The ratios of the masses of the quarks with the same electric charge and of the masses of the neutrinos
seem to be universal:

 \begin{equation}
 \begin{array}{c}
\frac{m^{}_u}{m^{}_c} \simeq \frac{m^{}_c}{m^{}_t} \simeq 0.005,\\
\\
\frac{m^{}_d}{m^{}_s} \simeq \frac{m^{}_s}{m^{}_b} \simeq 0.044,\\
\\
\frac{m^{}_1}{m^{}_2} \simeq \frac{m^{}_2}{m^{}_3} \simeq 0.25.
\end{array}
\end{equation}
\\
The dynamical reason for this universality is unclear. But the
mass ratios of the charged leptons are not universal:
\begin{eqnarray}
m(c) : m(u) \simeq m(t) : m(c) \simeq 207,\nonumber \\
m(s) : m(d) \simeq m(b) : m(s) \simeq 23.
\end{eqnarray}
If the ratio of the muon mass and the electron mass would be equal to the ratio of the tauon mass and the muon mass, the mass of the electron would have to be about ten times larger ($\simeq 6 ~MeV$).\\

Due to radiative corrections this universality is not expected to be exact.
A radiative correction of the order of $ \pm (\alpha/\pi){m^{}_\tau} \simeq \pm 4~MeV $ would have
to be added to the charged
lepton masses. Such a contribution is relatively small for the muon and the tauon, but it is large
for the electron. One expects that the physical electron mass
is the sum of a bare electron mass $\widetilde{m}_e$, due to the texture zero mass matrix, and a radiative
correction $r^{}_e$: ${m^{}_e} =\widetilde{m}_e - r^{}_e$. The ratio of the bare electron mass and the muon mass should be equal to the ratio of the muon
mass and the tauon mass: $\widetilde{m}_e$ $\simeq$ 5.51 MeV, $r^{}_e$ $\simeq$ 5.00 MeV. Radiative
corrections also contribute to the muon and the tauon mass, but
here the corrections are small in comparison to the bare masses and can be neglected.\\

We calculate the angle $\theta^{}_l $ (see eq.(15)) and obtain a result, which agrees with the experimental value:

 \begin{eqnarray}
\tan\theta^{}_l \simeq
\sqrt{\widetilde{m}_e/m^{}_\mu}\simeq 0.23, \nonumber \\
\theta^{}_{l}\simeq 13^\circ.
\end{eqnarray}
\\
In this paper we have shown that mass matrices with four "texture zeros" describe well the flavor mixing of quarks and leptons. The mixing angles for the
quarks and the leptons are functions of the fermion masses. The results agree with the experimental data.\\

\end{document}